\documentclass[grl]{agutex}
\usepackage{lineno}
\usepackage{graphicx}
\usepackage{amsmath}

\authorrunninghead{DAI ET AL.}
\titlerunninghead{Ion frozen-in condition in Reconnection}

\authoraddr{Lei Dai, Chi Wang, State Key Laboratory of Space Weather, Center for Space Science and Applied Research, Chinese Academy of Sciences, Beijing, China}
\authoraddr{Vassilis Angelopoulos, Department of Earth, Planetary and Space Sciences and Institute of Geophysics and Planetary Physics, University of California Los Angeles}
\authoraddr{Karl-Heinz Glassmeier, University of Braunschweig, Germany}

\begin{document}

\title{In situ Evidence of Breaking the Ion frozen-in Condition via the Non-gyrotropic Pressure Effect in Magnetic Reconnection}

\authors{Lei Dai\altaffilmark{1,2}, Chi Wang\altaffilmark{1}, Vassilis Angelopoulos\altaffilmark{3}, Karl-Heinz Glassmeier\altaffilmark{4}}

\altaffiltext{1}{State Key Laboratory of Space Weather, Center for Space Science and Applied Research, Chinese Academy of Sciences, Beijing, China.}
\altaffiltext{2}{School of Physics and Astronomy, University of Minnesota,Minneapolis, MN, USA}
\altaffiltext{3}{Department of Earth, Planetary and Space Sciences and Institute of Geophysics and Planetary Physics, University of California Los Angeles, Los Angeles, CA, USA.}
\altaffiltext{4}{University of Braunschweig, Germany}

\begin{abstract}

For magnetic reconnection to proceed, the frozen-in condition for both ion fluid and electron fluid in a localized diffusion region must be violated by inertial effects, thermal pressure effects, or inter-species collisions. It has been unclear which underlying effects unfreeze ion fluid in the diffusion region. By analyzing in-situ THEMIS spacecraft measurements at the dayside magnetopause, we present clear evidence that the off-diagonal components of the ion pressure tensor is mainly responsible for breaking the ion frozen-in condition in reconnection. The off-diagonal pressure tensor, which corresponds to a nongyrotropic pressure effect, is a fluid manifestation of ion demagnetization in the diffusion region. From the perspective of the ion momentum equation, the reported non-gyrotropic ion pressure tensor is a fundamental aspect in specifying the reconnection electric field that controls how quickly reconnection proceeds. \\

\end{abstract}

\begin{article}

\section{Introduction}
Magnetic reconnection is considered to drive global-scale dynamics in Earth's magnetosphere \citep{Angelopoulos2008Sci,Angelopoulos2013Sci}, solar flares \citep{Parker1957,Petschek1964}, laboratory plasmas and astrophysical systems. Magnetic reconnection changes the magnetic field topology and releases magnetic energy into particle energy in plasmas. As reconnection occurs, magnetic field lines `break' and `reconnect' at the X-line. Outside the diffusion region, plasma motions are frozen to magnetic field lines that behaves like elastic strings. Within the diffusion region, the violation of plasma frozen-in allows magnetic field lines to disconnect and reconnect.\\

The name `diffusion region' originates from the early Sweet-Parker reconnection model in which oppositely directed magnetic field lines (\textbf{B}) in the current sheet diffuse into the plasma ($\partial_t\textbf{B}\sim\nabla^2\textbf{B}$) as plasmas are demagnetized by inter-species collisions \citep{Parker1957,Petschek1964}. The term `diffusion region' has been extended to include any physical process that can violate the frozen-in condition \citep{Vasyliunas1975,Hesse1999}. The physics in the limited diffusion region is of high importance in magnetic reconnection. The global-scale evolution of the magnetic topology relies on the reconfiguration of magnetic field lines in the diffusion region. The mechanisms responsible for breaking the frozen-in condition specify the reconnection electric field that controls how quickly reconnection proceeds. \\ 

For reconnection to proceed, the frozen-in condition for both ion and electron must be violated in the diffusion region. In case that only one species is unfreezed from the field line, one can trace the frozen magnetic field lines tied to the other magnetized species. The exact condition for unfreezing ion fluid is a non-zero curl of $\textbf{E}+\textbf{v$_i$} \times \textbf{B}$ based on the frozen-in theorem. According to the ion fluid momentum equation without approximation, 

\begin{equation}
\textbf{E}+\textbf{v}_i \times \textbf{B}=(1/n_iq_i)\nabla \cdot \texttt{P}_i+(m_i/q_i)d\textbf{v}_i/dt+(m_i/q_i)\nu_{ie}(\textbf{v}_i-\textbf{v}_e),	
\end{equation}

where \textbf{E} is the electric field, $\texttt{P}_i$ is the ion pressure tensor, $\textbf{v}_i$ ($\textbf{v}_e$) is the ion (electron) flow velocity, and $\nu_{ie}$ is the effective ion-electron collision frequency. As evident from the ion momentum equation (1), the ion frozen-in condition can be violated by three non-ideal effects, i.e., an anisotropic ion pressure tensor, an ion inertial (acceleration) effect or inter-species collisions manifested as friction. These non-deal effects are the fluid manifestation of the kinetic effects leading to ion demagnetization in reconnection. The expression (1) and arguments for ions applies equally to electrons after a suitable change of charge, mass and sub-indexes.\\

The main non-ideal effects for ion and electron fluid in reconnection is still an open question. Reconnection models are categorized according to which non-ideal process is thought to be dominant. In the resistive reconnection model, inter-species collision is the dominant effect. The collisions can be either binary or anomalous (induced by wave-particle interactions that are not clearly understood). In collisionless reconnection models where effective ion-electron collisions are infrequent, the non-ideal effects have to be anisotropic pressure effects and/or inertial effects. As plasmas are convected toward the reconnection site, ions first become demagnetized at the characteristic scale length of ion motions while electrons are still magnetized \citep{Sonnerup1979book}. Electrons are expected to be demagnetized at a smaller spatial scale. \\

The reconnection electric field, defined in the co-moving frame of the current layer, controls how quickly (by $\textbf{E}\cdot$ \textbf{J}) magnetic energy is converted to particle energy in reconnection. The reconnection electric field is the same in the ion momentum equation (Eq.1) and the electron momentum equation. From the perspective of the ions (electrons), the reconnection electric field is the sum of the non-ideal electric field supported by ion (electron) non-ideal effects and the simple ion (electron) convection electric field. Although demagnetization is more difficult for electrons, the reconnection electric field is equally explainable in terms of ions or electrons near the X-line \citep{Hesse2011,Cai1997}. In this sense, ion demagnetization and electron demagnetization are two different but equally fundamental aspects in specifying the reconnection electric field.\\

In collisionless reconnection, ion decoupling from the magnetic field lines is considered to produce a Hall effect \citep{Sonnerup1979book,Nagai2001,Oieroset2001}. But what non-ideal effect is responsible for unfreezing ion fluid in the first place has not been clear. Intuitively, it is tempting to attribute ion decoupling to the Hall current term in the generalized Ohm's law, $\textbf{E}+\textbf{v$_i$}\times\textbf{B}=(1/nq)\textbf{J}\times\textbf{B}$. But this relation is a trivial equivalence of the electron frozen-in condition (\textbf{E}+\textbf{v$_e$}$\times$\textbf{B}=0). Ion information is canceled out from this relation and nothing specific about non-ideal effects of ions can be inferred. Studies of ion-scale physics in reconnection have been usually guided by the generalized Ohm's law, which is essentially a combination of electron and ion momentum equations with some approximations. Although appropriate for electrons to a certain extent, these approximations neglect and hide important aspects of ion dynamics. The correct approach to investigate ion scale physics is the full momentum equation without approximation \citep{Hesse1999,Hesse2011,Cai1997}. \\

Experimental clarification of the ion non-ideal effects requires a comparison between the non-ideal electric field (\textbf{E}+$\textbf{v$_i$} \times \textbf{B}$) and ion pressure/inertial terms. Although attempts have been made to compare the non-ideal reconnection electric field with the divergence of the electron pressure tensor , the electron diffusion region was too small for the spacecraft to encounter \citep{Henderson2006}. On the ion skin depth scale, past spacecraft observations have reported ion kinetic features such as non-gyrotropic ions and counterstreaming ions during reconnection \citep{Hoshino1998JGR,Nagai1998JGR,wygant2005,Zhou2009JGR,Aunai2011ANG}. But how these effects are linked to ion demagnetization and the violation of the ion frozen-in condition has been elusive. Here we report observations from an encounter of the THEMIS spacecraft \citep{Angelopoulos2008SSR,Sibeck2008SSR} with the diffusion region near the reconnection X-line at the Earth's magnetopause. The comprehensive fields instrumentation and measurements of ion velocity distribution \citep{Bonnell2008SSR,McFadden2008SSR} on THEMIS spacecraft provide an ideal opportunity to address the question of ion demagnetization in reconnection. By comparing the non-ideal reconnection electric field with ion pressure/inertial terms, we for the first time identify the non-ideal effects corresponding to ion demagnetization in the diffusion region. \\

\section{Data analysis}
On Feb 13, 2013, THEMIS spacecraft E moved into the magnetopause and detected a reconnection diffusion region. THEMIS E was at (6.1,-6.9,-0.6) in units of Re (Earth radii) in the Geocentric Solar Magnetospheric coordinate system (GSM). We adopted the boundary normal coordinate \textbf{NML} to study the magnetopause current sheet. \textbf{N} (-0.78,0.59,0.21)GSM is the boundary normal direction (outward) as determined by the minimal variance direction of magnetic fields, \textbf{L}=(-0.07, -0.42, 0.90) GSM is the direction of maximum variance of the magnetic field, and \textbf{M} completes the right-handed orthogonal coordinate, directed out of plane. Figure 2 shows measurements of fields and particles during the magnetopause crossing. The magnetopause current sheet, indicated by a change of \textbf{B}$_L$ from -40 nT to 60 nT, was observed from 23:24:40UT to 23:25:00UT. Northward plasma flow with velocities as large as 130 km/s was detected from 23:24:42UT to 23:25:05UT, immediately after the spacecraft crossed the separatrix S1 at 23:24:40UT. The separatrix S1 is identified by a sudden increase of 1-10 keV electrons that originated in the magnetosphere. The electron characteristics are good indication of separatrix field lines because their small gyroradius and large mobility along the magnetic field. Around 23:25:10UT, THEMIS E crossed the separatrix S2 and moved into the magnetosphere. The separatrix S2 is identified by a boundary between a broad spectrum of mixed electrons and a dominant magnetosphere electrons. Between 23:25:10UT and 23:25:30UT, spacecraft was away from the current sheet and the diffusion region, as evidenced by the large \textbf{B}$_L$ component. After 23:25:30UT, THEMIS E moved back into the magnetopause and observed the southward plasma flow. \textbf{B}$_N$ shows a variation of a few nT on a 20s timescale in addition to the DC component near the current sheet. This variation of \textbf{B}$_N$ may result from an eigenmode of the current sheet surface waves \citep{Dai2009,Dai2011}. The DC \textbf{B}$_N$ is negative in the northward plasma flow and positive in the southward flow, consistent with the prediction that the bidirectional plasma flows are accelerated by a \textbf{J}$\times$\textbf{B}$_N$ slingshot effect resulting from reconnection. The observed bidirectional reconnection outflows indicate that the spacecraft moved from the northward side to the southward side of an active reconnection X-line. \\

Figure 2g exhibits a significant deviation of the measured electric field \textbf{E}$_M$ from $-\textbf{v$_i$}\times\textbf{B}$ from 23:24:44UT to 23:24:58UT during the current sheet crossing. This deviation illustrates the violation of the ion frozen-in condition as a support of the reconnection electric field \textbf{E}$_M$, experimentally defining the so-called ion diffusion region. From 23:25:20UT to 23:25:44UT is the edge of the diffusion region and far from the current sheet center. The ion diffusion region in the current sheet is marked by a pink rectangle. Within the diffusion region, a remarkable normal electric field \textbf{E}$_N$ as large as 10 mV/m was observed. The out-of-plane magnetic field \textbf{B}$_M$ displayed one bump around the current sheet center (\textbf{B}$_L\sim0$); it was about 10 nT larger than the average guide field outside the current sheet. The location and waveforms of \textbf{E}$_N$ and \textbf{B}$_M$ are completely consistent with those of the Hall magnetic fields and electric fields in the asymmetric reconnection reported in early THEMIS observations \citep{Mozer2008GRL,Bonnell2008SSR}. Contrary to the situation of a quadrupole Hall \textbf{B}$_M$ and a bipolar Hall \textbf{E}$_N$ in a symmetric reconnection, there is only one bump in the Hall field in the asymmetric reconnection. The observed asymmetric Hall electric fields and magnetic fields indicate the operation of collisionless reconnection.  \\

The length of the diffusion region along the normal direction is the crossing time $\Delta$T multiplied by the magnetopause velocity v$_n$ relative to the spacecraft. We estimate v$_n$ by assuming that the magnetopause had a constant tangential electric field as it moved in the normal direction at constant speed. The estimated v$_n$ is [E$_M$(1)-E$_M$(2)]/[B$_L$[1]-B$_L$[2]] \citep{Mozer2002}. Here the fields E$_M$(1),E$_M$(2),B$_L$[1] and B$_L$[2] are measured at the spacecraft frame at times 1 and 2 upstream and downstream of the current sheet, respectively. As shown in Figure 2d and 2g, E$_M$(1) is $\sim$0.5mV/m and B$_L$[1] is$\sim-$45 nT from 23:24:20UT to 23:24:40UT before the crossing. Immediately after the crossing E$_M$(2) is $\sim$2.5mV/m and B$_L$[2] is$\sim$70 nT from 23:25:02UT to 23:25:14UT. The magnetopause velocity V$_n$ is $\sim$ 17.4 km/s. The length of the diffusion region along the normal direction is 264.5 km, about 7.3 magnetosheath ion skin depths or 1.2 magnetosphere ion skin depths ($c/\omega_{pi}$). This scale size is consistent with the values of the diffusion region in previously reported examples of collisionless reconnection \citep{Mozer2002,Vaivads2004,wygant2005}. The estimated tangential reconnection electric field near the reversal of \textbf{B}$_L$ is $\sim$0.5-1.3mV in the frame co-moving with the magnetopause current sheet. \\     

The departure of the pressure tensor from cylindrical symmetry about the local magnetic field direction can be measured by a nongyrotropy index \citep{Aunai2013}.The nongyrotropy and similar agyrotropy indexes have been successfully applied to characterize non-gyrotropic electrons in the electron diffusion region \citep{Scudder2012PRL,Tang2013GRL,Aunai2013}.
The comparison between nongyrotropy and non-ideal electric field, however, has not been done yet in these studies. In our event, Figure 2h shows an intensified layer of ion nongyrotropy within the diffusion region, indicating strongly non-gyrotropical ions when the ions are demagnetized. The ion nongyrotropy is as large as 0.3 in the diffusion region, significantly larger than the average value ($\sim$0.05) outside the diffusion region. Both the off-diagonal pressure component and the difference between the diagonal component can contribute to the nongyrotropy index \citep{Aunai2013}. We find that the contribution from the off-diagonal pressure component is dominant ($>$90\%) in the diffusion region. \\

The non-gyrotropic pressure can lead to the violation of the ion (or electron) frozen-in condition via the off-diagonal terms in the pressure tensor. As similar to the situation of non-gyrotropic electron pressure \citep{Vasyliunas1975,Hesse1999,Hesse2011}, the gradient of the off-diagonal ion pressure components can give rise to a non-ideal reconnection electric field in the diffusion region, (\textbf{E}$+\textbf{v$_i$} \times \textbf{B})_{M}\sim(\partial \texttt{P}_{NM}/\partial N+ \partial \texttt{P}_{ML}/\partial L)/n_iq_i$. Figure 2i presents the ion pressure tensor components in the boundary normal coordinate system. Spin-resolution ion moments data has been extensively used in the Walen test (or tangential component test) \citep{Sonnerup1981JGR}. Ion moments data are generally considered OK when the spacecraft takes multiple samples in the reconnection current layer \citep{Hoshino1998JGR,Nagai1998JGR,Oieroset2001,Mozer2002,wygant2005,Zhou2009JGR,Aunai2011ANG}. From 23:24:46UT to 23:24:54UT, THEMIS observed a significant non-ideal reconnection electric field (\textbf{E}$+\textbf{v$_i$} \times \textbf{B})_{M}$ of 3-4.5 mV/m (Figure 2g) in the center of the diffusion region associated with a gradient of the ion off-diagonal pressure component (Figure 2i). The density n$_i$ (Figure 2b) was $\sim$6.2cm$^{-3}$ at 23:24:50UT, $\Delta$N from 23:24:46UT to 23:24:54UT is $\sim$-129km, \texttt{P}$_{NM}$ (Figure 2i) decreases from 2500eVcm$^{-3}$ to near 0, $\Delta$\texttt{P}$_{NM}$ $\sim$ -2500eVcm$^{-3}$. With all these numbers, our estimate of $\Delta \texttt{P}_{NM}/\Delta Nn_iq_i$ is +3.2mV/m. $\Delta \texttt{P}_{ML}$ is $\sim$ -1000eVcm$^{-3}$ in Figure 2i. Assuming $\Delta$L$\sim$ $\Delta$N, $\Delta \texttt{P}_{ML}/\Delta Ln_iq_i$ is $\sim$ 1.3mV/m. The assumption of $\Delta$L$\sim$ $\Delta$N corresponds to an ion diffusion region extending 16 ion skin depth (roughly from 23:24:46UT to 23:25:20UT) northward in the \textbf{L} direction. Such \textbf{L}-extent of ion diffusion region is consistent with reconnection models. The gradient of the off-diagonal ion pressure terms agrees well with the non-ideal reconnection electric field, indicating that the non-gyrotropic pressure effect mainly contributed to breaking the ion frozen-in condition. Ion inertial (acceleration) effects and, in principle, the anomalous collision effect can also contribute to reconnection electric fields. In the diffusion region, V$_N$$\sim$-70km/s and $\Delta$V$_M$$\sim$130 km/s (Figure 2a). The ion inertial term scales as $(m_i/q_i)$V$_N$$\Delta$V$_M$/$\Delta$N ($<$0.3mV/m), much less than the non-gyrotropic pressure effect. The anomalous collision term is expected to be very small, because the observed reconnection should be collisionless as implied by current sheet thickness (ion skin depth) and the Hall fields. The ion pressure tensor can be decomposed into a gyrotropic part and a non-gyrotropic part. As shown in figure 2j, the non-gyrotropic part of the pressure tensor mostly contributes to the off-diagonal components. Based on the analysis of all possible effects in the ion momentum equation, we demonstrate that the non-gyrotropic pressure effect is primarily responsible for breaking the ion frozen-in condition in this event.\\

The ion momentum equation can be checked term by term in the normal direction as well. (\textbf{E}$+\textbf{v$_i$} \times \textbf{B})_{N}$ is about 7-10 mV/m near the reversal of \textbf{B}$_L$. The ion inertial term in the normal direction is on the order of 0.1 mV/m. The gradient of the pressure term in the normal direction is $(\partial \texttt{P}_{NN}/\partial N+ \partial \texttt{P}_{LN}/\partial L)/n_iq_i$. The $\Delta \texttt{P}_{NN}$ $\sim$ -6000eVcm$^{-3}$ is the dominant term, contributing to 7.8 mV/m in the ion momentum equation. Within a difference of 20\%, the pressure gradient term is approximately consistent with the (\textbf{E}$+\textbf{v$_i$}\times \textbf{B})_{N}$. This independent verification of the ion momentum equation in the normal direction involves the quantitative comparison between three independent data sets, indicating that THEMIS measurements were reasonably reliable in the ion diffusion region. \\  

As implied from the the two-fluid momentum equation (1), electrons are expected to exhibit similar dynamics as ions except that the non-ideal electron terms operate on the smaller electron scale in the diffusion region. Theoretical studies show that non-gyrotropic pressure effect unfreeze electron fluid near the reconnection X-line \citep{Hesse1999,Kuznetsova2001}. Encompassing the observation of ion non-gyrotropic pressure, an emerging scenario is that the non-gyrotropic pressure effect applies universally to unfrozen ions at large spatial scale and electrons at small spatial scale. Near the X-line, the reconnection electric field should be equally expressed in terms of ion or electron non-gyrotropic pressure \citep{Hesse2011}. \\

Figure 3 shows the ion velocity distributions perpendicular to the magnetic field near and in the diffusion region. Panel a shows a typical gyrotropic ion distribution in the magnetosheath outside the diffusion region. The distributions in panel b and c are centered around \textbf{B}$_L$=-20nT and \textbf{B}$_L$=+40nT, where the magnetic field are found to deviate from the spin-averaged values by $\sim$30\% and $\sim$20\% during one spin period, respectively. The velocity distributions in panels b and c are in the clearly identified ion diffusion region and characterized by non-gyrotropic bulges in the core component. According to the definition of pressure tensor $\texttt{P}=m\int dv^3(\textbf{v}-\textbf{u})(\textbf{v}-\textbf{u})f(\textbf{x},\textbf{v},t)$, where \textbf{u} is the bulk flow velocity, such non-gyrotropic bulges are expected to contribute to non-zero integral of $(\textbf{v}_M-\textbf{u}_M)(\textbf{v}_N-\textbf{u}_N)$ and thus large off-diagonal pressure components. Two bulges are centered at (50,15) km/s and (-82,-50) km/s in the interval c. The velocity spread of these bulges is $\sim$100km/s, about half of the thermal speed in the diffusion region. The observed non-gyrotropic distribution is particularly important for magnetic field line reconfiguration in the diffusion region as it can drive a current sheet instability that induces the normal magnetic field component\citep{Motschmann1998ASS}. In addition, the non-gyrotropic distributions are expected to drive plasma waves that tend to stabilize the plasma \citep{Motschmann1998}. The strongly non-Maxwellian distribution observed in the diffusion region has two implications: 1) the effective ion-electron collision is insufficient to lead to Maxwellian distributions while ions are demagnetized, and 2) the ion demagnetization process is much faster than ion thermalization and thus unlikely to be explained by effective ion-electron collisions. The non-gyrotropic bulges in panels b and c resemble those reported by GEOTAIL spacecraft in reconnection \citep{Hoshino1998JGR,Nagai1998JGR,wygant2005}. Numerical simulations suggest that non-gyrotropic bulges are caused by the mixing of already-accelerated meandering ions with ions just convected into the vicinity of the X-line geometry \citep{Hoshino1998JGR,Hoshino1998GMS}. Such a mixing process is inferred to also occur in the asymmetric reconnection at the magnetopause from observations from CLUSTER \citep{Lee2014JGR}. Ions from distinct sources form phase-bunched bulges in the distribution. This scenario of forming non-gyrotropic bulges is supported by the difference between the energy of the non-gyrotropic bulges shown in panels b and c. The nongyrotropy in the warmer ion components in panel d resemble those in THEMIS observations outside the current sheet center but near the reconnection site \citep{Zhou2009JGR}. These non-gyrotropic warmer ions are formed by a cucumber-type trajectory that has the neutral sheet-crossing and the non-crossing segments near an X-line geometry \citep{Zhou2009JGR}.\\  

\section{Summary and Conclusions}
Ion thermal pressure effects have been neglected and the ion inertial effect is difficult to identity in the generalized Ohm's law. Therefore, the individual ion momentum equation without approximation is the best choice for investigating the ion-scale physics in reconnection. We take this approach and present clear evidence that a non-gyrotropical ion pressure situation is responsible for breaking the ion frozen-in condition in reconnection. For the first time, we show that the non-ideal reconnection electric field associated with the violation of ion frozen-in condition is consistent with the gradient of the off-diagonal ion pressure components. Our approach and results would be valuable for the MMS (Magnetospheric Multiscale Mission) mission that is designed to resolve all scales relevant to magnetic reconnection.\\

Electrons are expected to exhibit similar dynamics as ions except that the non-ideal effect kicks in at smaller electron scale. An emerging scenario is that non-gyrotropic pressure effect universally unfreezes ions at ion scale and electrons at smaller scale in the collisionless reconnection. As evident from the two-fluid momentum equation, the reconnection electric field is equally explainable in terms of ions or electrons. Near the X-line, the reconnection electric field may be equally expressed in terms of ion or electron non-gyrotropic pressure effect \citep{Cai1997,Hesse2011}.  \\ 

\begin{acknowledgments}
THEMIS Data are available at \\
http:$//$themis.ssl.berkeley.edu. We thank the THEMIS and ancillary data providers. L.D. thanks Steve Monson for proofreading the manuscript. We thank Judy Hohl for editorial help. K.-H.G was financially supported through the German Ministry for Economy and Energy and the German Center for Aviation and Space (DLR) under contract 50 OC 0302. This work was supported by NASA contract NAS5-02099, NNSFC grant 41231067 and in part by the Specialized Research Fund for State Key Laboratories of China. \\
\end{acknowledgments}

\bibliographystyle{agu08}

%
%

\end{article}

\clearpage

\begin{figure}
\noindent
\includegraphics[width=35pc]{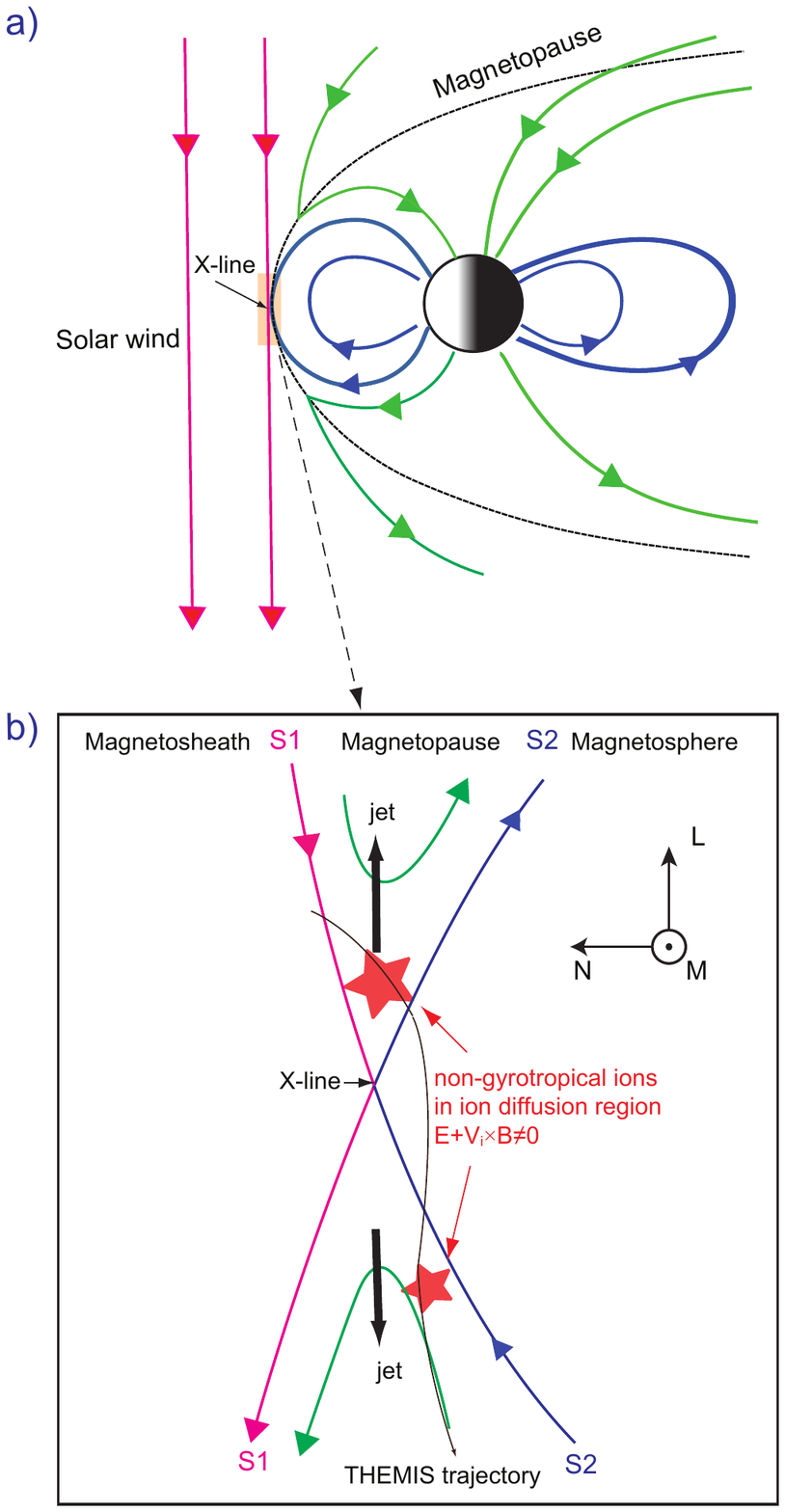}
\caption{Schematics of magnetic reconnection in the Earth's magnetopause. \textbf{a}, The magnetic field geometry at the dayside magnetopause reconnection viewed in the noon-midnight plane. Magnetic field lines near magnetopause reconnection can be divided into three classes according to their topology : (1) interplanetary field lines (red) with no magnetic foot on the earth, (2)`open' field lines (green) with one magnetic foot connected to the earth and (3) `closed' field lines (blue) with both magnetic feet connected to the Earth. Two branches of separatrix surface S1 and S2 intersect along a magnetic `X-line' directed out of the plane. \textbf{b}, The expanded view of the region surrounding the X-line. A portion of the plasma outflow came across the ion diffusion region identified by a significant deviation of \textbf{E} from $-\textbf{V$_i$} \times \textbf{B}$. }
\label{Fig1}
\end{figure}

\begin{figure}
\noindent
\includegraphics[width=35pc]{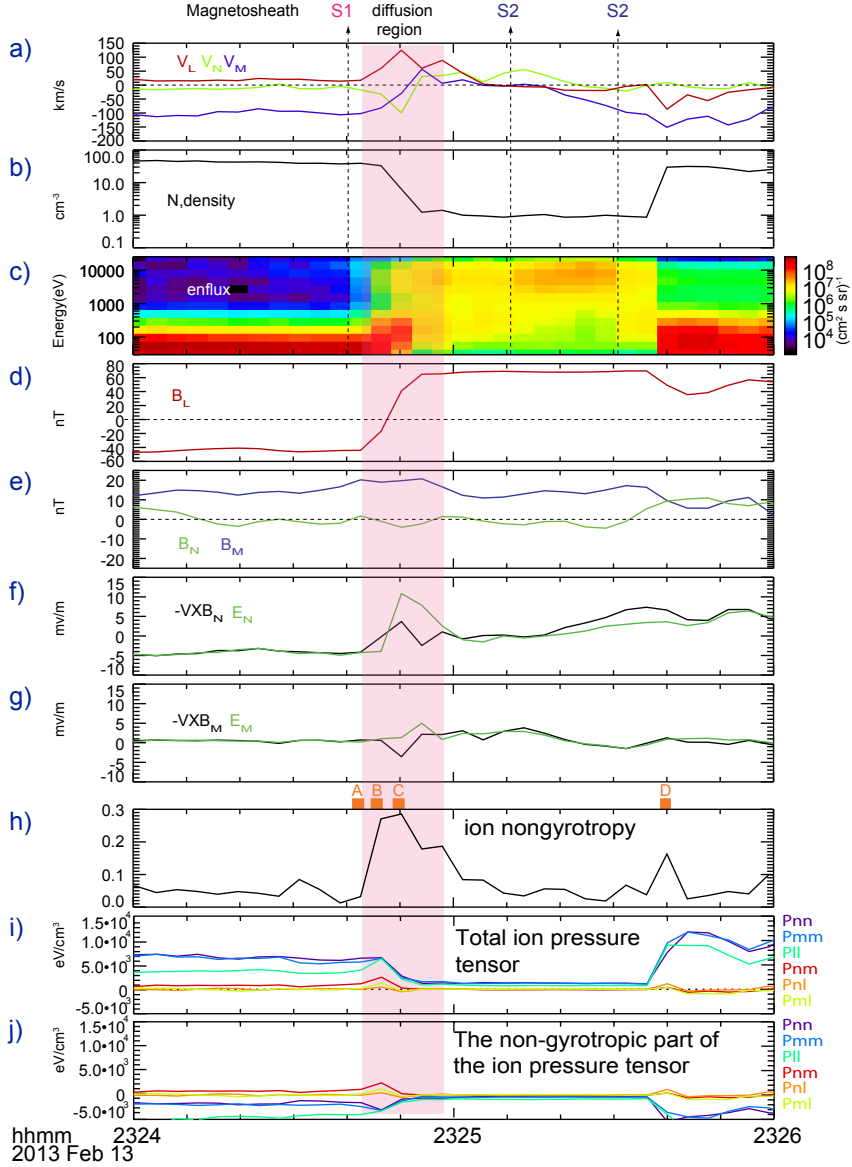}
\caption{Observations by THEMIS spacecraft E in the ion diffusion region. All THEMIS data in Figure 2 are of spin-resolution and at the cadence of the particle instrument. \textbf{a}, Three components of the proton flow velocity in the \textbf{LMN} coordinate system, showing a flow reversal in the \textbf{L} (North-south direction). \textbf{b}), The plasmas density. \textbf{c}, The electron differential energy flux. \textbf{d-e}, three component of magnetic field in the \textbf{LMN} coordinate. \textbf{f,g}, The comparison of $\textbf{v$_i$} \times \textbf{B}$ with the electric fields in \textbf{N} and \textbf{M}, respectively. Three dimensional electric fields are obtained from the \textbf{E} $\cdot$ \textbf{B}=0 assumption. \textbf{h} The ion nongyrotropy index \citep{Aunai2013}, characterizing the degree of nongyrotropy in the distribution function. The ion velocity distribution function in intervals marked by A,B,C and D are show in Figure 3. \textbf{i}, Components of the ion pressure tensor in the boundary normal coordinate. \textbf{j}, The non-gyrotropic part of the ion pressure tensor.}
\label{Fig2}
\end{figure}

\begin{figure}
\noindent
\includegraphics[width=35pc]{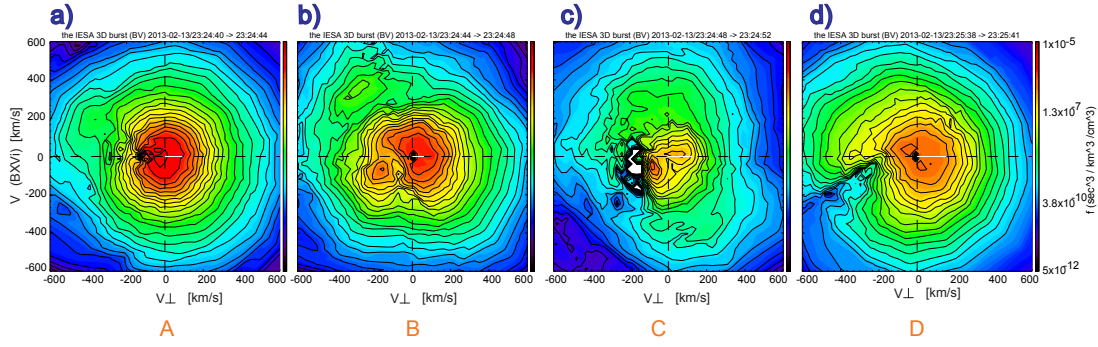}
\caption{Observations of ion velocity distributions near and within the ion diffusion region. \textbf{a-d}, The cut of velocity distribution functions perpendicular to the magnetic field measured from THEMIS ESA in the intervals marked as `A,B,C,D' respectively in Figure 2. The horizontal axis is the direction of ion convection flow \textbf{V}$_{i\bot}$ and the vertical axis is the direction of \textbf{B}$\times$\textbf{V}$_{i\bot}$. The ion bulk flow velocity has been subtracted from the distributions. The white horizontal line in the distribution represents the flow convection velocity in the spacecraft frame. Interval a is far from the diffusion region, b and c are within the diffusion region and d is at the edge of the current layer and away from the current sheet center.}
\label{Fig3}
\end{figure}

%
%

\end{document}